\documentclass[traditabstract]{aa} 
\def\taud{\tau_\mathrm{d}}

\def\javelin{$\mathit{JAVELIN}$}
\def\gp{\mathcal{GP}}
\newcommand{\bea}{\begin{eqnarray}}
\newcommand{\eea}{\end{eqnarray}}
\newcommand{\be}{\begin{equation}}
\newcommand{\ee}{\end{equation}}
\newcommand{\bc}{\begin{center}}
\newcommand{\ec}{\end{center}}
\newcommand{\ben}{\begin{enumerate}}
\newcommand{\een}{\end{enumerate}}
\newcommand{\bd}{\begin{description}}
\newcommand{\ed}{\end{description}}
\newcommand{\bmi}[1]{\begin{minipage}{#1 cm}}
\newcommand{\emi}{\end{minipage}}
\newcommand{\bmif}[1]{\begin{minipage}{#1\textwidth}}

\def\rund#1{\left( #1 \right)}

\def\runddd#1{\left\{ #1 \right\}}

\def\av#1{\bar{#1}}

\def\Real{{\rm I\mathchoice{\kern-0.70mm}{\kern-0.70mm}{\kern-0.65mm}%
  {\kern-0.50mm}R}}
\def\C{\rm C\kern-.42em\vrule width.03em height.58em depth-.02em
       \kern.4em}

\def\bx#1{\leavevmode\thinspace\hbox{\vrule\vtop{\vbox{\hrule\kern1pt
        \hbox{\vphantom{\tt/}\thinspace{\bf#1}\thinspace}}
      \kern1pt\hrule}\vrule}\thinspace}

{\catcode`\@=11
\gdef\SchlangeUnter#1#2{\lower2pt\vbox{\baselineskip 0pt \lineskip0pt
  \ialign{$\m@th#1\hfil##\hfil$\crcr#2\crcr\sim\crcr}}}
}

\def\ueber#1#2{{\setbox0=\hbox{$#1$}%
  \setbox1=\hbox to\wd0{\hss$\scriptscriptstyle #2$\hss}%
  \offinterlineskip
  \vbox{\box1\kern0.4mm\box0}}{}}

\usepackage{graphicx}

\usepackage{color}

\usepackage{txfonts}
\usepackage{rotating}
\usepackage{multirow}
\usepackage{url}
\usepackage{txfonts}
\usepackage{natbib}
\usepackage{subfigure}
\usepackage{enumerate}

\def\Hbeta{\ion{H}{$\beta$}}

\def\MgII{\ion{Mg}{II}}

\def\CIII{\ion{C}{III]}}
\def\CIV{\ion{C}{IV}}

\begin{document}
   \title{ Imprints of the quasar structure in time-delay light curves: Microlensing-aided reverberation mapping}
 \author{D. Sluse
 \inst{1} 
 \and M. Tewes
 \inst{1}
 }

 \institute{{Argelander-Institut f\"ur Astronomie, Auf dem H\"ugel 71, 53121 Bonn, Germany.}
 \email{dsluse@astro.uni-bonn.de}
 }

\date{Received ; accepted}

  \abstract
{Owing to the advent of large area photometric surveys, the possibility to use broad band photometric data, instead of spectra, to measure the size of the broad line region of active galactic nuclei, has raised a large interest. We describe here a new method using time-delay lensed quasars where one or several images are affected by microlensing due to stars in the lensing galaxy. Because microlensing decreases (or increases) the flux of the continuum compared to the broad line region, it changes the contrast between these two emission components. We show that this effect can be used to effectively disentangle the intrinsic variability of those two regions, offering the opportunity to perform reverberation mapping based on single band photometric data. Based on simulated light curves generated using a damped random walk model of quasar variability, we show that measurement of the size of the broad line region can be achieved using this method, provided one spectrum has been obtained independently during the monitoring. This method is complementary to photometric reverberation mapping and could also be extended to multi-band data. Because the effect described above produces a variability pattern in difference light curves between pairs of lensed images which is correlated with the time-lagged continuum variability, it can potentially produce systematic errors in measurement of time delays between pairs of lensed images. Simple simulations indicate that time-delay measurement techniques which use a sufficiently flexible model for the extrinsic variability are not affected by this effect and produce accurate time delays. }

\titlerunning{Imprints of the quasar structure in time-delay light curves}
\authorrunning{D. Sluse et M. Tewes}


   \maketitle

\section{Introduction}

Gravitational lensing of distant quasars offers new opportunities in the study of quasars and super massive black holes. When a galaxy happen to be on the line of sight towards a more distant quasar, multiple images of the latter are created, with typical separation of a few arcsecs. The (macro-) magnification of the lensed images associated to this phenomenon ease the study of those distant objects and allow one to reconstruct a high resolution image of the quasar's host \citep{Peng2006b, Claeskens2006, Suyu2013a}. Since the stars in the lensing galaxy also act as many individual gravitational lenses, the macro images are in fact composed of many unresolved micro-images. The latter are separated on the sky by a few micro-arcsec corresponding the Einstein radius $\eta_0$ of the microlenses. The microlensing effect can be used to probe the source at high resolution because it selectively magnifies its emission as a function of the size of the emitting region, provided the latter is smaller than $\sim 10\,\eta_0$ \citep{Refsdal1997}. Nowadays, quasar microlensing is employed to measure the size and temperature profile of the accretion disc, or the size and geometry of the broad line emitting region \citep{Kochanek2004a, Sluse2007, Eigenbrod2008b, Morgan2010a, Blackburne2011a, Sluse2011a, Odowd2011, Guerras2013}. 

The quasar continuum emitting region is more compact than $\eta_0$, and is therefore significantly microlensed. The more extended broad line region (BLR) is generally less affected, with typically 10-20\% of its flux being microlensed \citep{Sluse2012b}. Consequently, microlensing effectively modifies the contrast between the flux of the continuum and the one from the broad lines. Because a lensed system is composed of several (two to four) lensed images of the quasar, we observe multiple realizations of the same intrinsic light curve with different amount of microlensing of the continuum and the broad lines. The proposed technique of {\it {microlensing-aided reverberation mapping}} aims at taking advantage of this effect to measure the time lag between the continuum and broad line variations. Our method is not conceptually very different from photometric reverberation mapping \citep{Haas2011, Chelouche2012, Chelouche2013, PozoNunez2012, Edri2012, Zu2014, Bachev2014}. In that case, multi-band photometry is used to disentangle the flux of the continuum and of the BLR, while our technique can already be applied to single-band data. The same data can now be used to derive the size of the continuum emission based on the variability of the microlensing signal \citep{Kochanek2004a}, and if multi-bands data are available, to measure the temperature profile of the accretion disc \citep{Anguita2008a}. This opens the possibility to study the properties of the accretion disc and of the BLR in the same population of AGNs. 

In Sect.~\ref{sec:fiducial}, we present a fiducial example of lensed quasar light curves which demonstrates that imprints from the continuum and broad line variations are present in the difference light curve between pairs of lensed images under the simplifying assumption of non-variable microlensing. We explain how we proceed to simulate daily sampled lensed quasar light curves, and show how to use them to measure the time lag between the continuum and the broad line variations. We also discuss how our results depend on the properties of the source, and show that this signal should be detected in time-delay light curves. In Sect.~\ref{sec:simu}, we increase the complexity of the simulations, producing mock light curves with irregular sampling and seasonal gaps, as well as microlensing signal drawn from microlensing simulations. We propose a technique to detrend the time variable microlensing signal with a B-spline model in order to recover the time lag. In the next section (Sect.~\ref{sec:dt}) we search for a bias on the measurement of the time delay in mock light curves which include continuum and BLR flux. Finally, Sect.~\ref{sec:conclusions} summarizes our main results and outlines future work necessary to turn this technique into a robust probe of the quasar structure at intermediate and high redshift. 

\section{Fiducial case and time-lag measurement}
\label{sec:fiducial}

In this section, we explain how we simulate realistic light curves of lensed quasars, and test simple techniques to measure the time lag between the continuum and the response from the BLR. We start with the ideal situation where a lensed quasar has been observed for 9 years on a daily basis. The duration of this mock light curve is chosen to mimic existing light curves of lensed quasars as obtained by the COSMOGRAIL collaboration \citep{Courbin2011, Tewes2013}, and light curves provided by future surveys like the large synoptic survey telescope (LSST). We first assume a noise-free light curve, and a constant amount of microlensing. Section~\ref{sec:simu} is dedicated to the simulation of more realistic light curves. Finally, for simplicity, we assume that the time delay from the pair of lensed images has been obtained independently and corrected for, or is naturally close to zero. The latter situation occurs for lensed systems where the four lensed images have a cross-like configuration around the lensing galaxy, e.g. the Einstein Cross Q2237$+$0305 \citep{Huchra1985}, or where two (resp. three) of the four lensed images are ``merging'', a situation happening when the source is located close to a ``fold'' (resp. ``cusp'') caustic, e.g. WFI~2033$-$4723 and RXS~J1131$-$1231 \citep{Morgan2004, Sluse2003}. Preliminary investigation of the impact of the presence of multiple sources of emission on time-delay measurements is discussed in Sect.~\ref{sec:dt}.

\subsection{Intrinsic variability}
\label{subsec:model}

First, we describe how we simulate the intrinsic variability of the lensed quasar. Despite our limited understanding of the detailed processes governing quasar variability, it has been shown by several authors \citep[e.g.][]{Kelly2009, Zu2013a, Graham2014} that Gaussian processes, and in particular damped random walk (DRW), provides a satisfying mathematical description of the AGN variability. Deviations from this model on time scales smaller than 5 days, as well as possible turnover in the properties of the signal for time scales above 54 days have been suggested~\citep[][]{Mushotzky2011, Zu2013a, Graham2014}. However, those deviations are relatively subtle and overall the DRW process provides a good proxy of the AGN variability. In the following, we generate mock AGN light curves using the \javelin \,code \citep{Zu2011,Zu2013a, Zu2014}. 

The continuum variability is described by: 
\begin{equation}
c(t) = \gp \runddd{\bar{c}, \kappa \rund{t, t^{\prime}}},
\label{eq:cont}
\end{equation}

\noindent where the mean function of the DRW is $\bar{c}$ (constant over time), and its associated covariance function between two epochs $t$ and $t^{\prime}$ is $\kappa \rund{t, t^{\prime}}$. Following several authors, we use an exponential covariance function of the form  $\kappa\rund{t, t'}=\sigma^2\exp\rund{-|t-t'|/\taud}$ where $\sigma^2$ and $\taud$ are the variance and characteristic time scale of the process \citep{Zu2014, Graham2014}. Note that instead of $\sigma$, various variability studies \citep{McLeod2010, Butler2011} of AGNs use $\hat \sigma$ which is the amplitude of the DRW. It is related to $\sigma$ through the relation $\sigma^2 = 0.5\,\tau \hat \sigma^2$ \citep{Kozlowski2010}. On long time scales, the variance of the light curve is $\hat \sigma (\taud/2)^{1/2}$ and on short time scales $\hat \sigma \sqrt{t}$. 

The variations in the broad line region is modeled as the variation of the continuum convolved with a time-lagging transfer function $\Psi(t)$:
\begin{equation}
l(t) = \int \Psi(t-t')c(t)\,dt', 
\label{eq:line}
\end{equation} 

Following \citep{Zu2011, Zu2014}, we use a top-hat transfer function centered on the time lag $\tau$, with width $w$ and amplitude $A$,
so that
\begin{equation}
    \Psi(t)\equiv \Psi(t|\tau, A, w) =  A / w \;\;\mbox{for}\;\; \tau - w/2 \leqslant t < \tau + w/2.
    \label{eq:tophat}
\end{equation}

\cite{Chelouche2012, Chelouche2013} suggest that photometric reverberation mapping using time-lag measurements based on cross-correlation methods are sensitive to the choice of $\Psi(t)$. However, the transfer function is observationally poorly constrained, as reverberation mapping studies generally concentrate on the measurement of the time-lag $\av\tau$ between the continuum and broad line variations, but not on recovering $\Psi(t)$. Since this work is a proof-of-concept of the technique, we will only use $\Psi(t)$ as defined in (\ref{eq:tophat}), which corresponds to a thin shell geometry of the BLR \citep{Peterson1993, Pancoast2011}. On the other hand, we also choose the properties of the intrinsic variability light curve to match qualitatively the variability of the lensed quasar RXS~J1131$-$1231 \citep{Sluse2003, Sluse2006, Tewes2013}. Future works should investigate in more details if results are biased by those particular choices. 

In the following, the total signal is assumed to be the sum of the continuum and of a single emission line. Furthermore, we fix $\tau=$100\,days. Since the bulk of time lags observed for \Hbeta\,in local AGNs is observed in the range 10-100\,days \citep{Kaspi2000, Bentz2009}, a lag of 100\,days may be representative of expected (cosmologically dilated) lags for high ionization UV lines in lensed AGNs at intermediate redshifts.

\subsection{Simulating the light curves}
\label{subsec:lcves}

The light curves of a pair of lensed images (at a given wavelength and to a good approximation in a given band; assuming that differential extinction due to the lensing galaxy is negligible, as commonly observed) can be expressed as: 

\begin{equation}
\begin{array}{l}
F_1(t) = M\,\mu_1(t)\,F_{M\mu}(t) + M F_M(t),\\
\vspace{2mm} F_2(t) = \mu_2(t)\,F_{M\mu}(t) + F_M(t), 
\end{array}
\label{eq:vary}
\end{equation}

\noindent where $F_{1,2}$ is the flux of the pair of lensed images, $M$ is the relative macro-magnification ratio between the pair of images{\footnote{If $M_{1,2}$ are the macro-magnification of images $1$ and $2$, then $M=M_1/M_2$.}}, $\mu_1(t)$ (resp. $\mu_2(t)$) is the amplitude of microlensing of the most compact region (i.e. the continuum) in image $1$ (resp. $2$), and $F_{M\mu}(t)$ ($F_{M}(t)$) the part of the flux which is (not) affected by microlensing. For the fiducial case, all the microlensed flux is emitted by the continuum as defined in \ref{eq:cont}, hence $F_{M\mu}(t) = c(t)$. On the other hand, the flux which is not microlensed originates from the broad line only, such that $F_{M}(t)=f_{BLR} \,l(t)$, with $f_{BLR}$ being the flux ratio between the line and the continuum. We should emphasize that in known lensed quasars, a small fraction (in general 10-20\%) of the flux from the BLR is microlensed \citep[see ][]{Sluse2012b}. Therefore, the above identification of the microlensed flux $F_{M\mu}$ to the continuum (only) and of the non-microlensed flux to the BLR is an approximation whose impact is discussed in Sect. \ref{subsec:lag}. In \ref{eq:vary}, and in the following, we assume that the macro-magnification ratio $M=M_1/M_2$, which corresponds to the continuum flux ratio which would be measured in absence of microlensing, has been derived independently from e.g. flux ratios measured from spectroscopy in narrow emission lines, or at longer wavelengths such as mid-infrared (MIR) or radio wavelengths (but see Sluse et al. \citeyear{Sluse2013} regarding presence of microlensing in the MIR). For simplicity, we set $M=1$. 

\begin{figure}  
\includegraphics[width=\hsize]{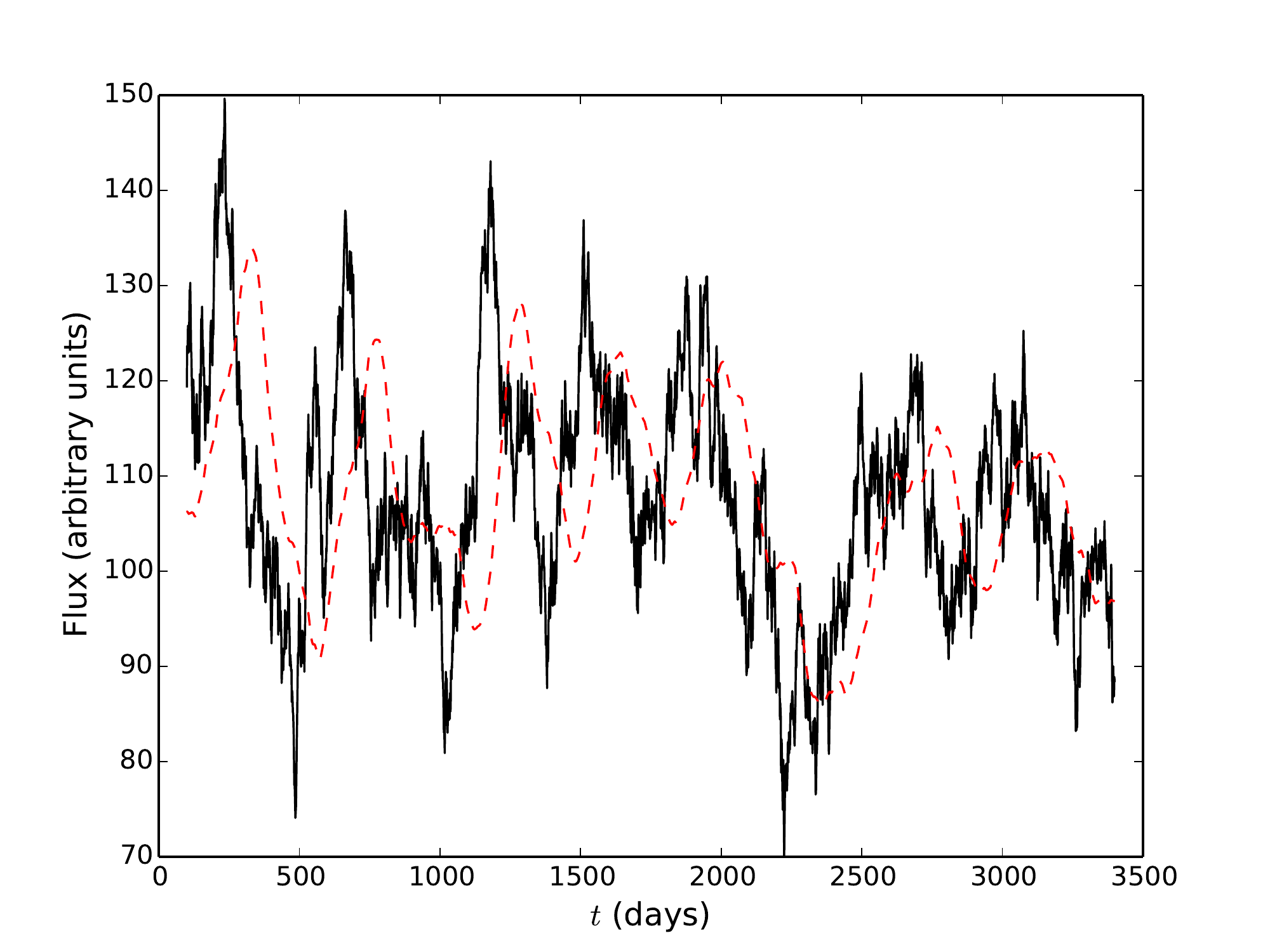}
\includegraphics[width=\hsize]{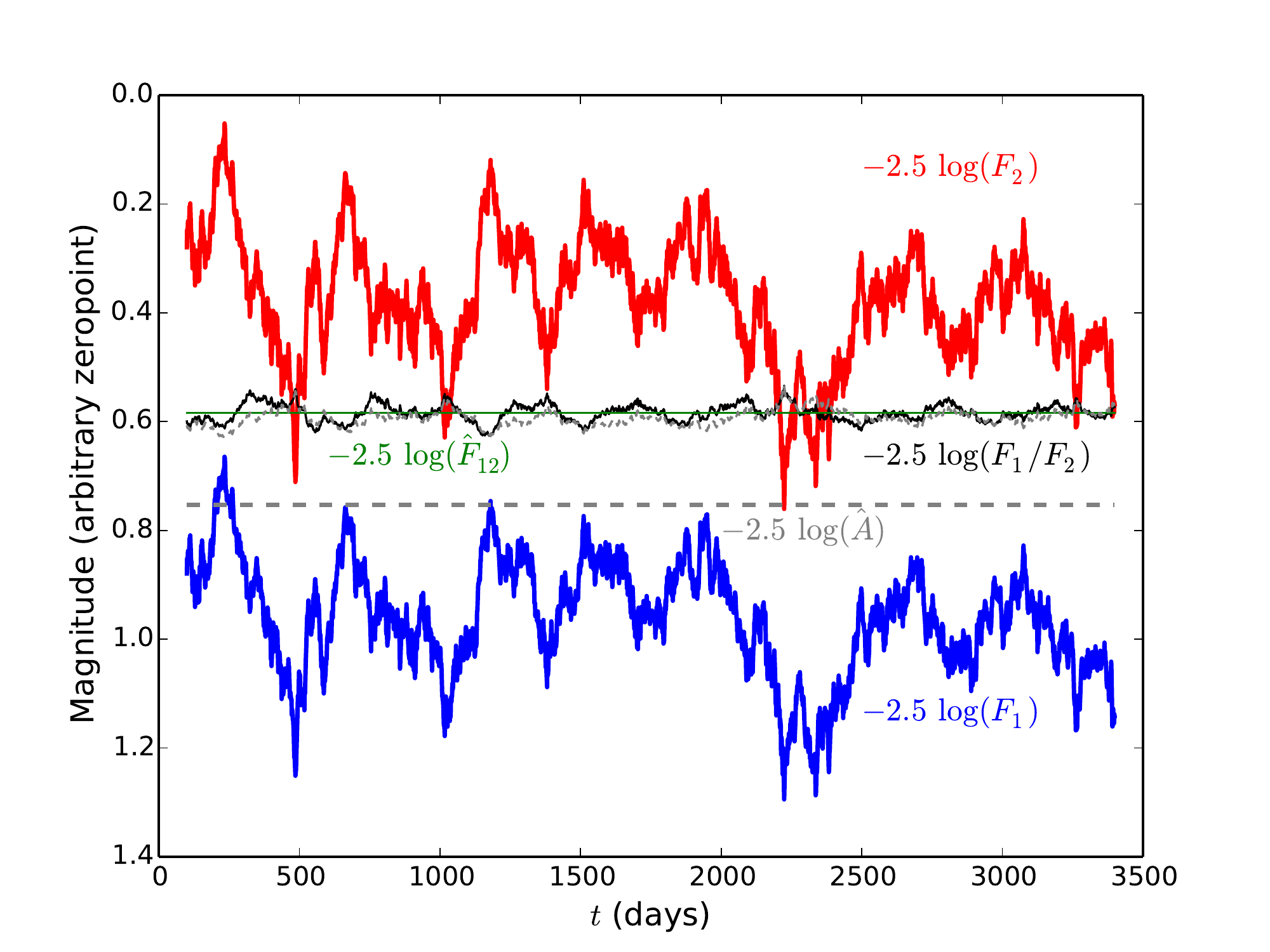}
\caption{{\it {Upper panel}}: Fiducial realization of the continuum (black) and emission line (red) light curves of a quasar (Sect.~\ref{sec:fiducial}). The light curve corresponding to the emission line has been divided by $f_{BLR}$ to show more clearly the time lag $\tau \sim 100$\,days between the continuum and the line. A negligible amount of noise has been considered to ease legibility. {\it {Bottom panel}}: Corresponding light curves of the two lensed images in magnitude (thick blue and red lines) and the corresponding difference light curve (solid black). The dotted gray line shows the differential light curve which would be observed if the broad line was not varying, the solid green line a model $\hat{F}_{12}$ of the extrinsic variations, and the thick dashed gray line is $\hat{A}(t)$.}

\label{fig:fidu}
\end{figure}

We start by using $f_{BLR}=0.2$. Although arbitrary, this choice may be representative of a large population of quasars. Indeed, the median equivalent width of the main UV emission lines (\MgII, \CIII, \CIV) is of the order of 40$\AA$ \citep{Croom2002, Shen2011} while the width of red optical filters is typically 120\,$\AA$. One may also note that some planned surveys will use narrow band filters \citep{Benitez2014, Marti2014} which would significantly increase $f_{BLR}$. In addition, we assume a constant microlensing in image $1$, $\mu_1(t)=0.5$ at all epoch, and no microlensing in image $2$. Simulations including more realistic microlensing signal are presented in Sect.~\ref{subsec:micro}. Figure~\ref{fig:fidu} shows the the continuum and emission line light curves in the top panel, and the simulated light curves of the pair of images in the bottom panel. The latter panel also shows in magnitude the differential light curve between images $1$ and $2$ (solid black line), and the same signal if only the continuum was varying (dashed line). Two important messages have to be kept from this figure. First, the presence of a fraction of the flux (time variable or not) which is not microlensed, leads to differential light curve which does not only contain signal from microlensing, as commonly assumed, but which is modulated at a detectable level (modulation of about $\pm$ 0.05 mag in Fig.~\ref{fig:fidu}) by a signal associated to the quasar intrinsic variability. Second, the differential light curve shows a different shape for a variable and a non-variable BLR. In the next section, we cross correlate the differential signal with the intrinsic signal in order to unveil the imprint of the time-lagged signal from BLR. 

\subsection{Cross Correlation Function}
\label{subsec:CCF}

\begin{figure}  
\includegraphics[width=\hsize]{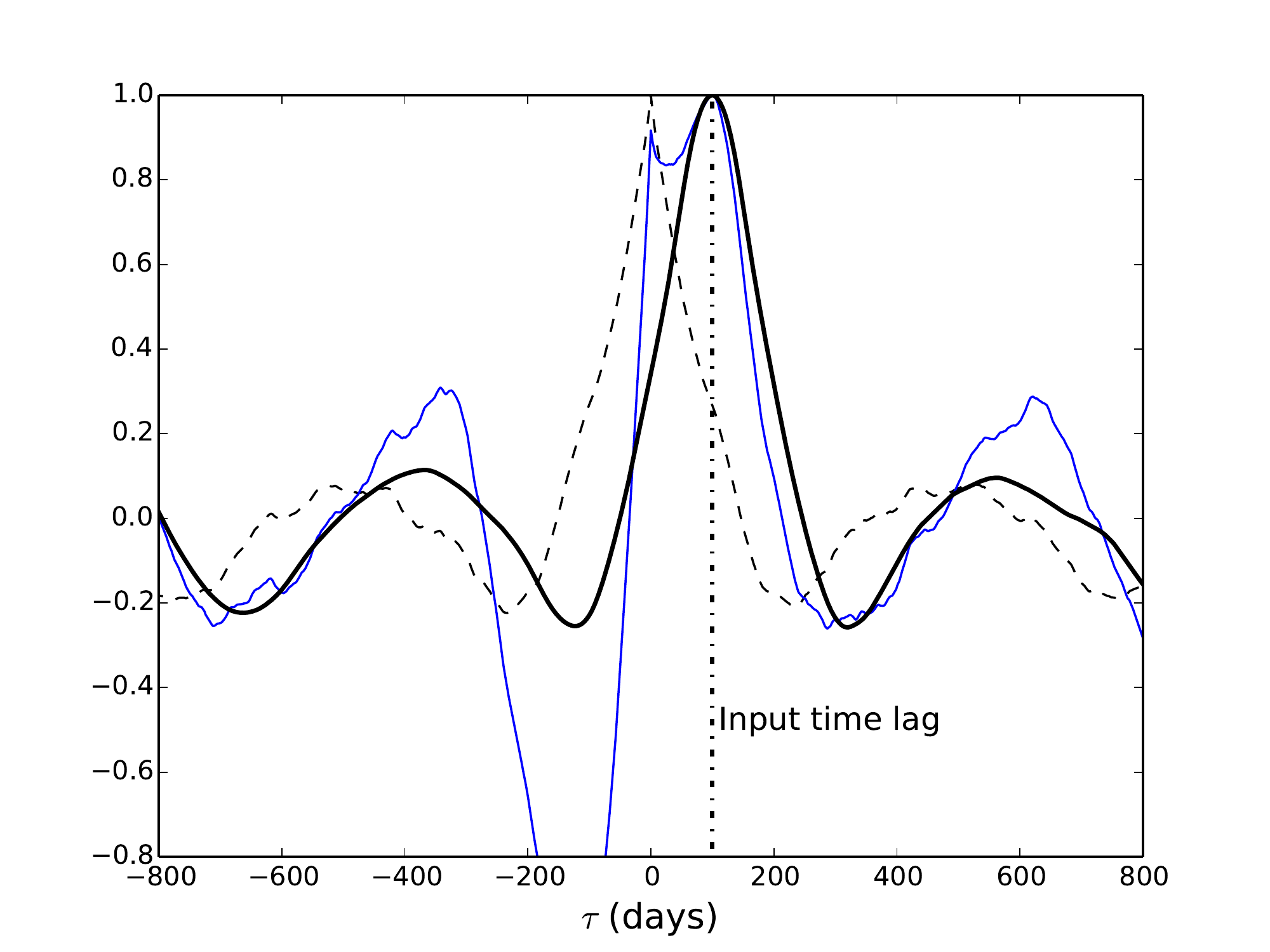}
\caption{Cross Correlation Function (normalized to peak intensity) of $F_2/F_1$ with $F_1$ in the case of a time-lagged BLR (thin blue line) and of a non-variable BLR (dashed black). The third curve (thick solid black line) shows the CCF between the fraction of the quasar flux affected by (free of) microlensing $F_{M\mu}$ ($F_M$), as obtained by linearly combining $F_1$ and $F_2$ (Sect.~\ref{subsec:MmD}, Equ. \ref{eq:MmD}). The dotted-dashed vertical line shows the fiducial time lag $\tau=$100\,days between the variation of the continuum and of the BLR in the simulation. }

\label{fig:CCF}
\end{figure}

Figure~\ref{fig:fidu} demonstrates that the difference light curve between two lensed quasar images is not the same when the contribution associated to the broad line region is constant or responds to the continuum variations. The cross correlation function (CCF) of the ratio light curve $F_2/F_1$ with the microlensed signal $F_1$ of the lensed image{\footnote{We choose $F_1$ as a reference because it maximizes the flux from the BLR while the differential signal $F_2/F_1$ contains mostly flux from the continuum. This choice is dictated by our knowledge of the fiducial signal.}} displayed in Fig.~\ref{fig:CCF}, confirms that the variations of the BLR are imprinted in the differential signal. The CCF nicely peaks at 100 days, namely the input time lag $\tau$, but it shows also a pronounced peak at 0 days, as the CCF obtained for a non-variable emission line. In fact, if one calculates the CCF between $F_2/F_1$ and $F_1$, this secondary peak becomes the main one. We discuss in Sect.~\ref{subsec:MmD} how linear combination of $F_1$ and $F_2$ can be used to reduce the power at a zero day lag, and more robustly retrieve the peak at the time lag $\tau$. 

\subsection{Macro-micro Decomposition (MmD)}
\label{subsec:MmD}

We describe here a method to deblend the continuum and the BLR signal from the intrinsic light curves. This technique is inspired by a similar method devised in Sluse et al. (\citeyear[][see also Hutsem\'ekers et al. \citeyear{Hutsemekers2010} and Sluse et al. \citeyear{Sluse2012b}]{Sluse2007}) but applied to quasar spectra instead of time series. Simple linear combinations of the signal of a pair of images, as expressed in \ref{eq:vary}, allows us to isolate $F_{M}$ using the observed fluxes $F_1$ and $F_2$: 

\begin{equation}
\begin{array}{l}
F_M(t) = \frac{-A(t)}{A(t)-M}\left(\frac{F_1(t)}{A(t)}-F_2(t)\right), \\
\vspace{2mm} \mu_2(t)\,F_{M \mu}(t) = \frac{M}{A(t)-M}\left(\frac{F_1(t)}{M}-F_2(t)\right), 
\end{array}
\label{eq:MmD}
\end{equation}

\noindent where we have introduced $A(t) = M \times \mu(t)$ (with $\mu(t) = \mu_1(t)/\mu_2(t) \ne 1$). This quantity is preferred to $M$ and $\mu(t)$ because it is more closely related to observations, with $A(t) = F_1(t)/F_2(t)$ when there is no flux from the emission lines. 

In order to perform this decomposition, it is necessary to know $M$ and to have a proxy to $A(t)$. The acquisition of a spatially resolved spectrum of the lensed images at an epoch $t_1$ (ideally part of the photometric monitoring period), with at least the same wavelength range as the broad band data, allows one to derive those two quantities. First, $M$ is obtained based on the flux ratio of the narrow emission lines, or of the broad lines if the latter are at least partly unaffected by microlensing \citep{Sluse2007, Hutsemekers2010, Sluse2012b, Braibant2014, Nierenberg2014}. Second, the flux ratio measured in the continuum of the spectra gives us $A(t_1)$. Third, it is possible to model $F_1(t)/F_2(t)$ with a smooth model $\hat{F}_{12} (t)$ which encodes the large scale {\it {extrinsic}} variations of $F_1(t)/F_2(t)$. For the example depicted in Fig.~\ref{fig:fidu}, such a model is the solid horizontal green line. We can then define an empirical estimate of $A(t)$ such that: 

\begin{equation}
\hat{A}(t)  =   \frac{A(t_1)}{\hat{F}_{12}(t_1)} \, \hat{F}_{12}(t).
\label{eq:A}
\end{equation}

\noindent In general $\hat{A}(t) \neq A(t)$, but the difference may not be large as far as the flux from the continuum is much larger than the flux from the BLR, and microlensing variations remain modest over the time of the monitoring. This is further discussed in Sect.~\ref{subsec:micro}. 

In summary, the MmD allows one to empirically deblend the signal which is microlensed, and mostly originating from the continuum emission, from the signal which is not affected by microlensing. The cross correlation of those two signals, is used in the following to measure the time lag $\tau$. Alternative methodologies may be developed, but we focus in this paper on the use of a zero-padded CCF applied to $F_M$ and $F_{M\mu}$ as derived with the MmD. 

\subsection{Modification of microlensing and BLR contributions}

Several properties of the signal might hamper the detectability of a lag, such as the relative contribution of flux from the line, $f_{BLR}$, the amount of microlensing from the continuum, $\mu$, and large photometric errors. In order to test those effects, we have simulated light curves in the same way as our fiducial light curves (i.e. time delay of 0 days, regular sampling of 1 point per day) for nine different values of $f_{BLR}$ uniformly distributed in the range $[0.1, 0.9]$, and for six values of amplitudes $\mu$, chosen such that $-2.5\,log(\mu)$ uniformly covers the range [-0.75, 0.75]\,mag ($\mu = 1$ excluded). First, we consider a noiseless situation. For each couple ($f_{BLR}$, $\mu$), we have generated 500 different light curves, and for each one we have measured the time lag using the peak of the CCF between $F_M$ and $F_{M\mu}$ as described in Sect.~\ref{subsec:MmD}. Following this procedure we retrieve a median time lag $\tau \sim 99$\,days, with a standard deviation $\sigma_{\tau} \sim 2.2$ days. The median time lag is 1\% smaller than 100 days because the distribution of time lags is asymmetric. The peak of the distribution is in fact found at 100 days. We have not identified the reason of this asymmetry. The transfer function does not seem to be the reason, as we obtained exactly the same average time lag when using a delta-function for $\Psi(t)$ in ~(\ref{eq:line}). A possibility could be low frequency variations of the quasar, known to produce similar biases in CCF analysis \citep{Chelouche2012}. Because time-lag measurements will be affected by larger errors than this bias at the percent level, we do not investigate it further as it will produce only a second order effect. 

\begin{figure*}
\begin{tabular}{cc}
\includegraphics[scale=0.45]{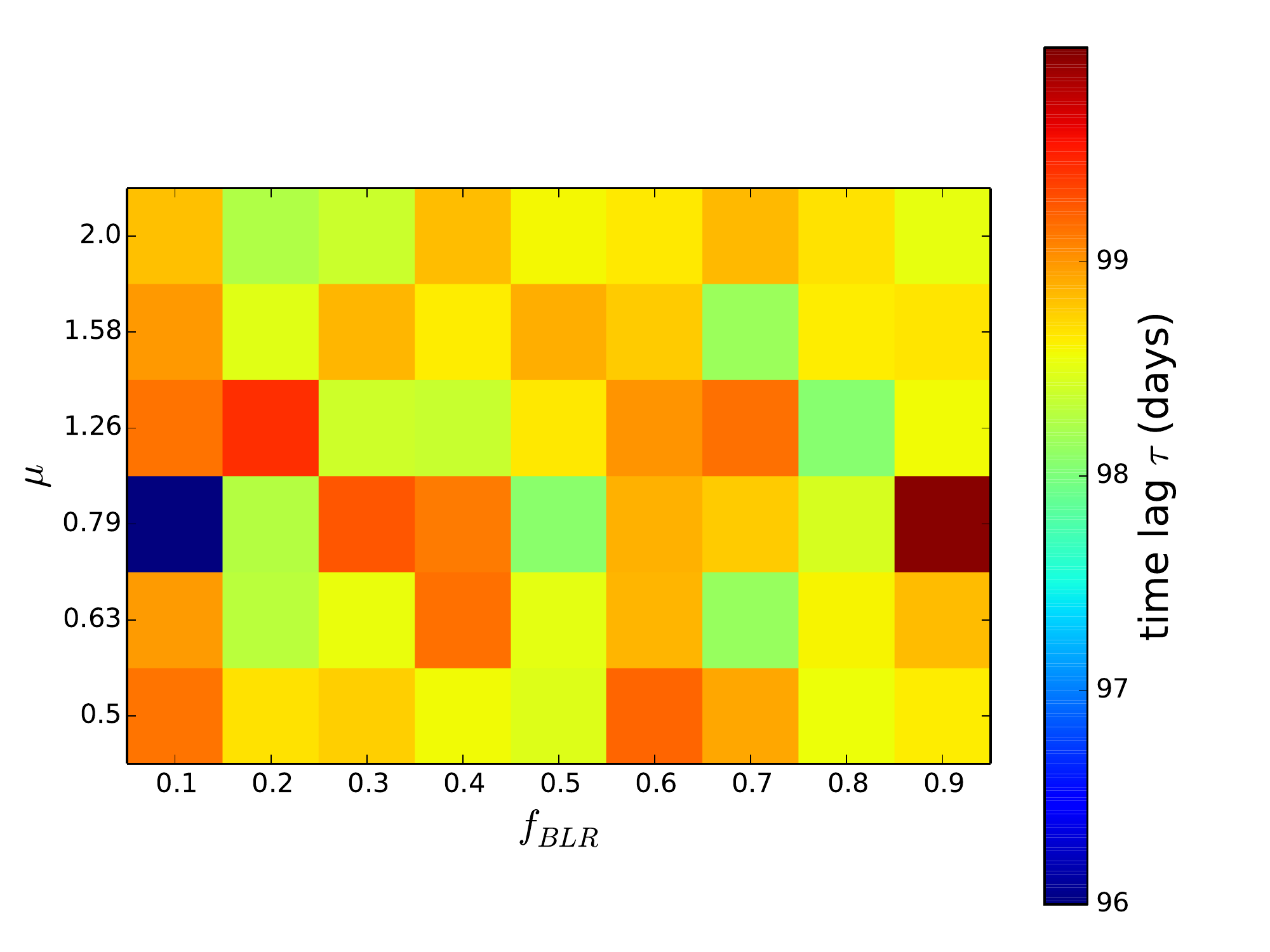} & \includegraphics[scale=0.45]{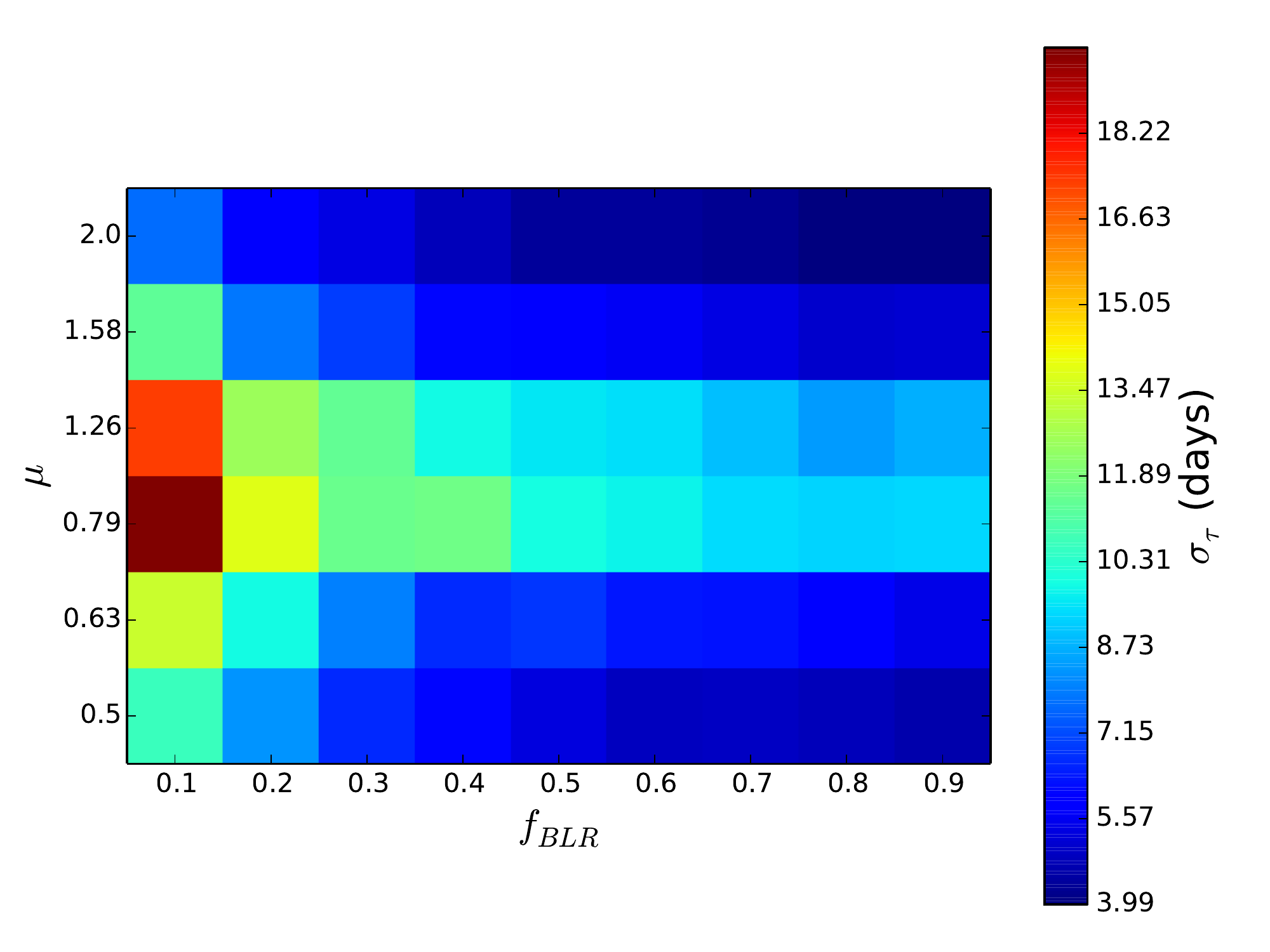} \\
\end{tabular}
\caption{Average time-lag measurement ({\it{left}}) and associated standard deviation ({\it {right}}) for various fractional flux contribution from the line $f_{BLR}$, and 6 different amplitude of microlensing of the continuum $\mu$. For each pair ($f_{BLR}$, $\mu$), 500 different quasar light curves have been generated, with a time lag $\tau = 100$\,days between the variation of the continuum and of the BLR. Note that due to the definition of $f_{BLR}$, the case $f_{BLR}=0.9$ corresponds to an average contribution of the line of 0.9/1.9 = 0.47 to the total flux. }
\label{fig:gridBLR}
\end{figure*}

Finally, we generate a second set of simulations, identical to the above ones, but to which we have added uniform Gaussian noise $\sigma_f/f = 0.01$. Such a small photometric uncertainty is aimed by upcoming surveys and is reached for good time-delay light curves currently obtained by monitoring programs such as COSMOGRAIL. Figure~\ref{fig:gridBLR} shows the results of this procedure. At this noise level, the mean time lag agrees perfectly with the time lag retrieved by noiseless simulations. Uncertainties (at 68.2 \% confidence level) smaller than 10 days can be obtained. However, as expected, the distribution of time lags broadens significantly (as revealed by the larger $\sigma_\tau$) when the amplitude of microlensing is small (i.e. $-2.5\,log(\mu) \sim \pm 0.25$\,mag) or for low values of $f_{BLR}$ (i.e., 0.1). The impact of the photometric accuracy on the results is discussed in the next section.

\subsection{Photometric accuracy}
\label{subsec:noise}

\begin{figure}
\includegraphics[scale=0.45]{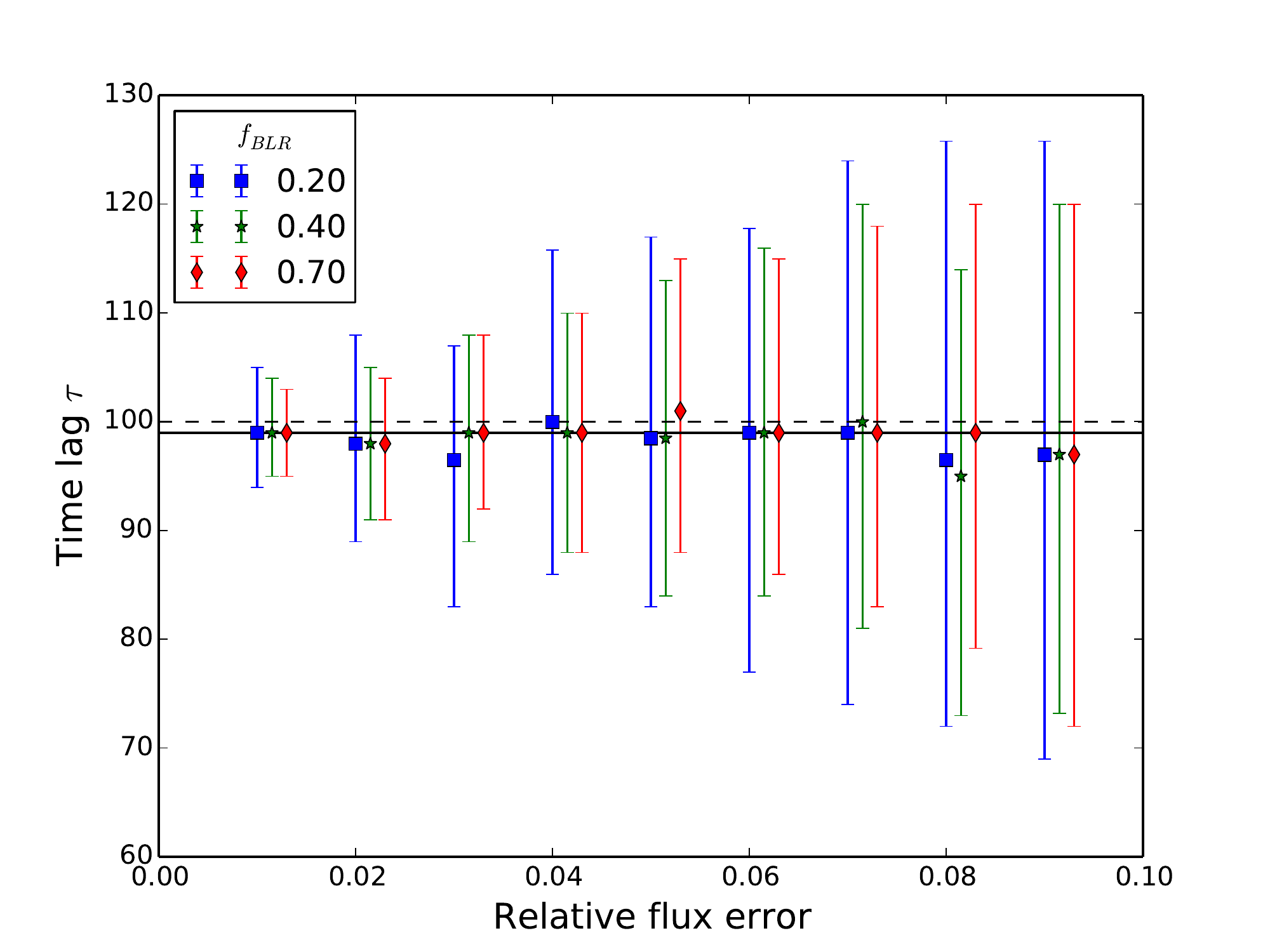}
\includegraphics[scale=0.45]{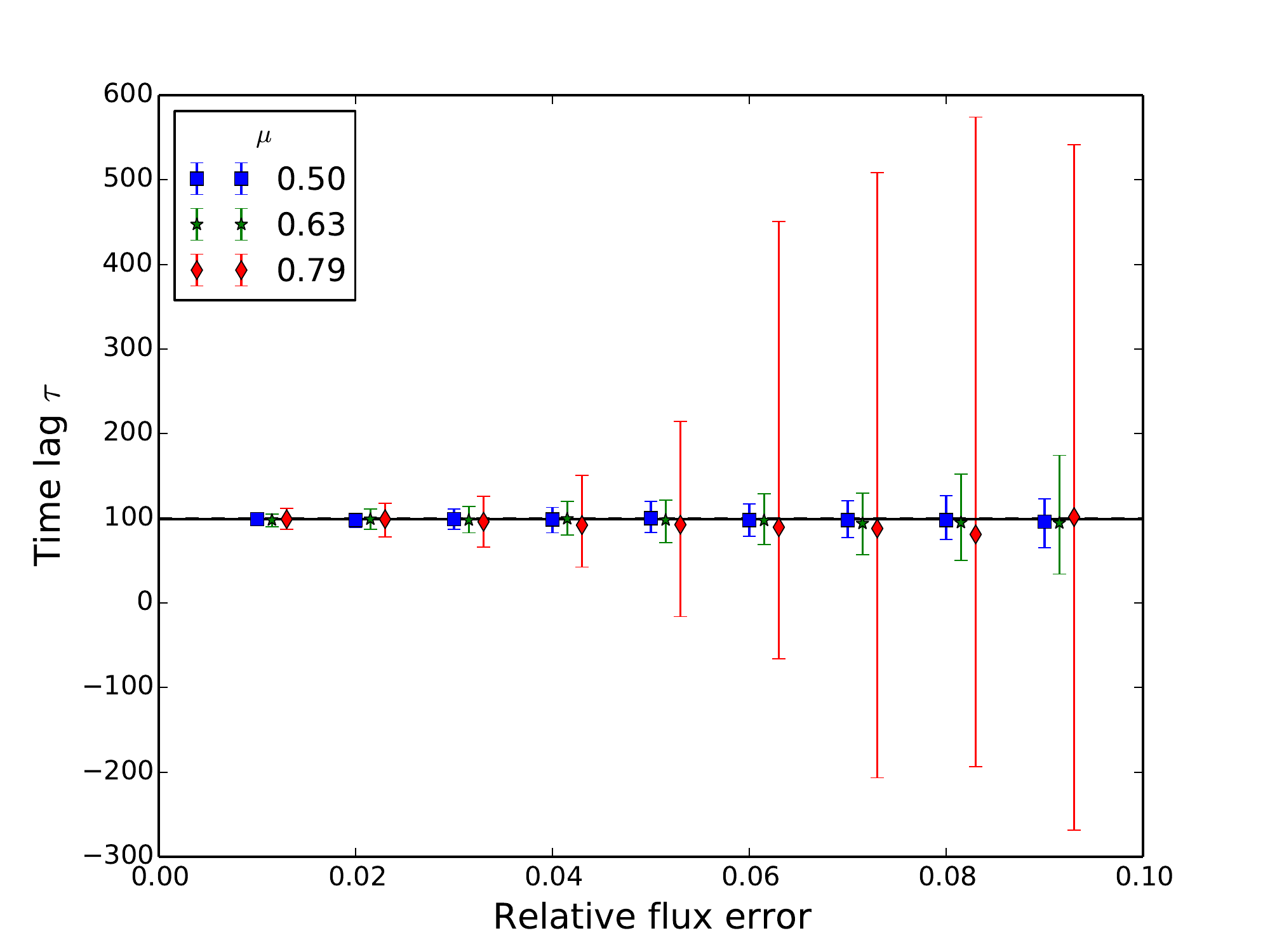}
\caption{Impact of the photometric errors (x-axis) on the distribution of retrieved time-lag measurements. {\it{Upper panel}}: $\mu = 0.5$. Three different values of $f_{BLR}$ are shown as stars, squares and diamonds. {\it{Lower panel}}: $f_{BLR} = 0.2$. Three different values of $\mu$ are shown as stars, squares and diamonds. The dashed black line shows the input time lag $\tau=100\,$days and the solid black line the median lag measured on light curves free of noise. The points with the same photometric uncertainty (i.e. stars, squares and diamonds) have been slightly shifted between each other towards larger uncertainty to ease legibility. In both panels, the solid blue squares correspond to our fiducial case for different photometric accuracies.   }
\label{fig:noisevar}
\end{figure}

The signal to noise of the input light curves is expected to limit the ability to measure $\tau$. For various pairs ($f_{BLR}$, $\mu$), we generate 500 noiseless simulated light curves. Then, we generate realizations of those light curves for different amounts of noise $\sigma_f/f \in [0.01, 0.09]$, and measure the time lag as described earlier. As shown above, in such a situation, the uncertainty in the results depends on the amplitude of microlensing $\mu$ and on the relative contribution of flux from the emission line, $f_{BLR}$. We study two cases: i) we fix $f_{BLR} = 0.2$, and vary $\mu=0.5, 0.63, 0.79$; ii) we fix $\mu=0.5$ and vary the amplitudes of $f_{BLR}= 0.2, 0.4, 0.7$. The median value of $\tau$, and $1\sigma$ interval (i.e. interval containing 68.2\% of the distribution) are displayed in Figure~\ref{fig:noisevar} as a function of the noise. Three conclusions can be drawn from that figure. First, the median of the distribution agrees with the noiseless case, and is biased low by 1.5\% compared to the true time lag, due to the asymmetry of the lag distribution. Second, for an amplitude of microlensing $\mu=0.5$, corresponding to a demagnification of the continuum by 0.75 mag, the uncertainty on the time lag increases by typically a factor 4 when the photometric uncertainty is increased from 0.01\,mag to 0.09\,mag. Third, the ability to measure a time lag at low signal to noise strongly depends on the amplitude of microlensing. While a time lag can still be measured in most of the cases for $(\mu, f_{BLR}) = (0.79, 0.2)$ if the photometric accuracy is better than 0.03 mag, this measurement becomes highly uncertain for lower signal to noise. We should emphasize that these trends should also depend on the amplitude of intrinsic variability of the quasar (and to some extent of the macro magnification $M$), and therefore are only representative of the variability properties assumed for our fiducial quasar.

\section{Simulations of more realistic light curves}
\label{sec:simu}

In the previous section, we have demonstrated that light curves of multiply-imaged quasars can be used, in presence of microlensing, to perform reverberation measurement of the size of the BLR. The fiducial light curves we presented are however highly idealized. Real light curves will be sampled on a less regular baseline and with gaps between seasons, while microlensing will not be constant but vary with time. In addition, a small amount of microlensing of the emission line may be expected. We consider the impact of all these features hereafter. This may however not cover all possible complication encountered in nature, and future work is needed to quantify how the method behaves when multiple emission lines fall in the same broad-band filter, and or when emission which is not time variable (e.g. host galaxy flux, narrow emission lines) is present.

\subsection{Sampling and gaps in the light curve}
\label{subsec:sampling}

Simulating sparsely sampled light curves with seasonal gaps can be performed in a simple way. For a given mock light curve, we first create a new light curve with regular gaps every year to mimic an ensemble of observing seasons, and then we reduce the sampling of each season by keeping a given fraction $f$ of the points per season. In practice, we create light curves with gaps of 130 days and keep only $f=34\%$ of the daily sampled points which corresponds to a mean frequency of observation of 1 point every 3 days. This kind of sampling is representative of the best light curves currently obtained for lensed quasars \citep{Vuissoz2008, Courbin2011, Tewes2013}. 

\begin{figure}
\includegraphics[scale=0.45]{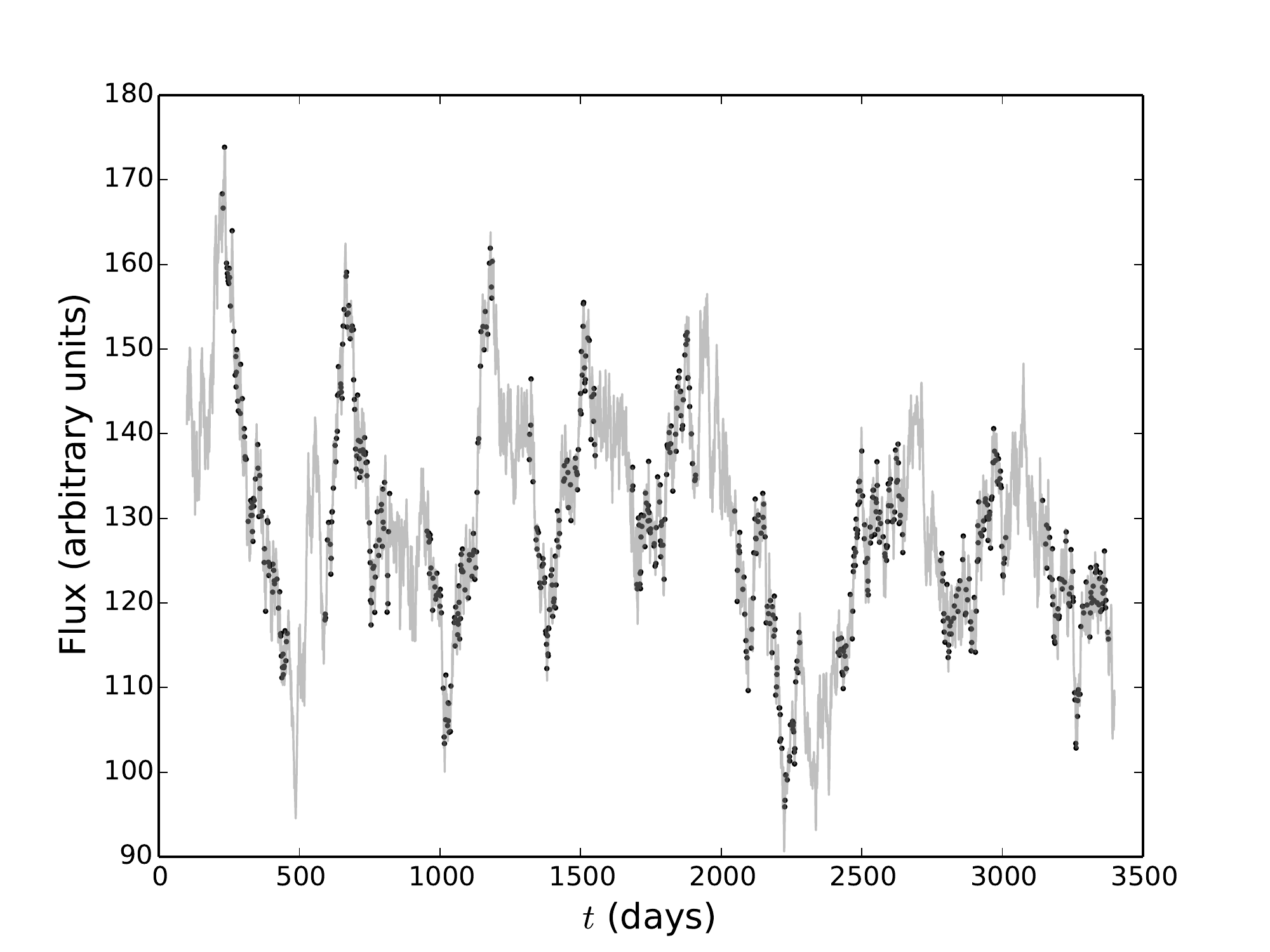}
\includegraphics[scale=0.45]{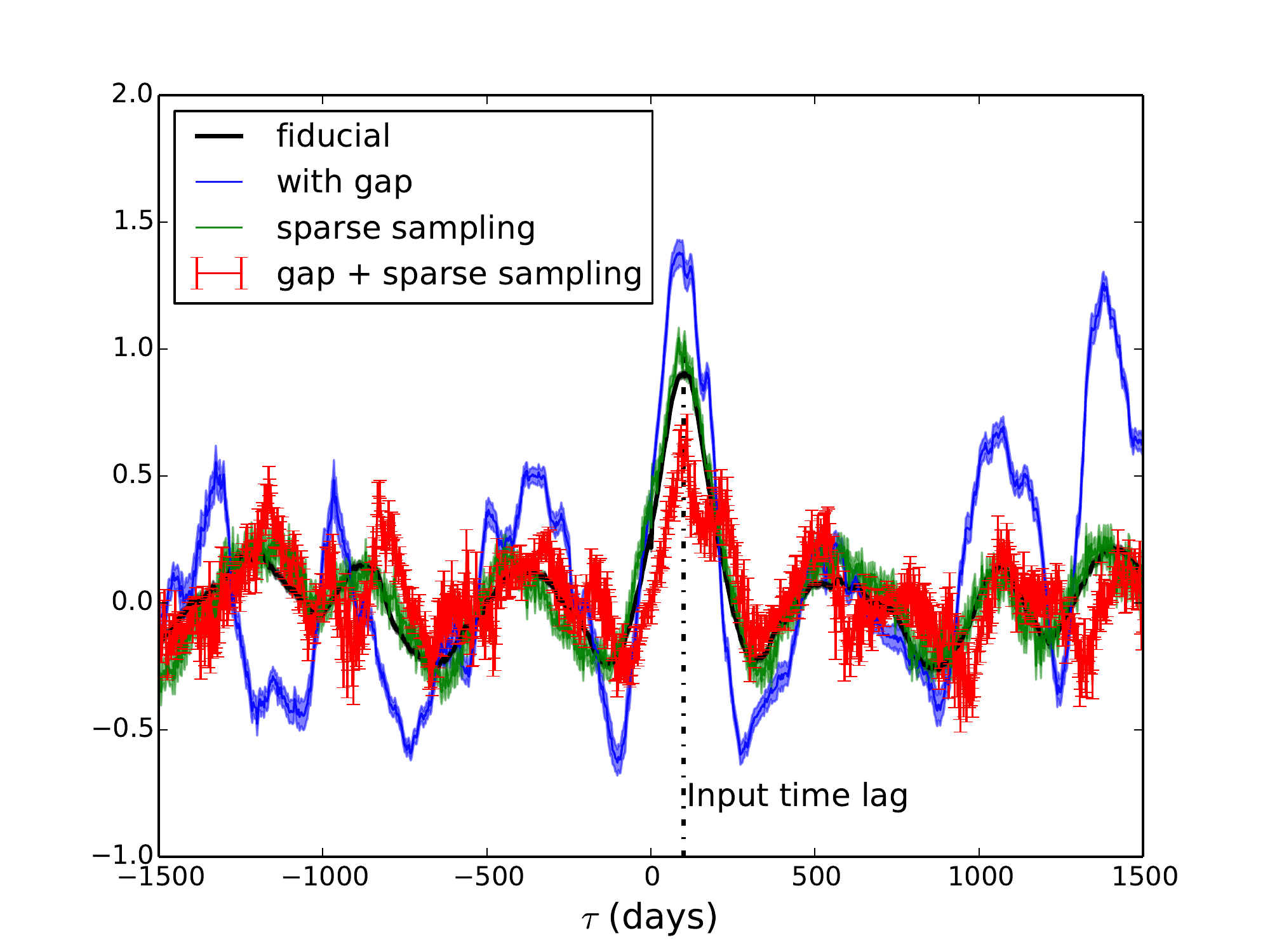}
\caption{Impact of gaps and sampling on the cross correlation. {\it{Upper panel:}} Realistic light curve (black circles) with irregular sampling and seasonal observing gaps, generated based on the fiducial ``continuous'' light curve (light solid gray). {\it{Bottom panel:}} Discrete Correlation Function (DCF) obtained for the fiducial ``continuous'' light curve (black band), when data are obtained with a mean sampling of 1 point every 3 days (green band), when data are obtained with a seasonal gaps of 130 days (blue band), and with a sparse sampling and seasonal gap as shown on the upper panel (red points with error bars).} 
\label{fig:DCFreal}
\end{figure}

The main complication when one works with sparsely sampled light curves comes from the use of the CCF which cannot be applied to irregularly sampled time series. This problem is common to all reverberation mapping measurements and several techniques have been introduced to address it. We use hereafter the most simple ones, namely the Interpolated Cross Correlation Function \citep[ICCF][]{Gaskell1987}, which consists in applying the cross correlation to the data set after interpolation, and the Discrete Correlation Function (DCF) introduced by \cite{Edelson1988}. An example of sparsely sampled light curve is shown in Fig.~\ref{fig:DCFreal}, with the corresponding cross correlation, calculated with a bin of 3 days (i.e. average seasonal sampling of the light curves) for the DCF. The DCF is shown for i) the fiducial ``continuous'' light curve, ii) with a sampling rate of 1 point every 3 days, iii) with a daily sampling but seasonal gaps, and iv) in case of sparse sampling and seasonal gaps. A fifth case, with the same number of points as (iv) but no gaps (i.e. a sampling rate of about 1 point every 5 days) has also been tested but is not shown as the DCF is similar to (ii). In all the cases, the amplitude of the correlation function is oscillating around zero, with a main peak around the true time lag. The DCF obtained with only one third of the points is not very different from the fiducial DCF, showing that the irregular sampling has little impact on the time lag measurement for observing rates of the order of a few days. More critical is however the presence of gaps in the light curves which lead to peaks with distorted shapes. Once light curves with both irregular sampling and seasonal gaps are considered, a secondary peak appears superimposed to the main one, at $\tau \sim $ 220 days. Interestingly, this secondary peak is not visible if we calculate the ICCF instead of the DCF. As we show hereafter, this secondary peak has important consequences for measuring the time lag when seasonal gaps are present. 

Instead of generating a large sample of different realizations of the continuum variability $c(t)$, we have repeated the procedure outlined above with different gap locations and different sampling of the input fiducial light curve. We have used the maximum of the correlation function to estimate the time lag. This quantity is well defined for the CCF studied until now, but is more sensitive to noise fluctuations when we calculate the DCF. Therefore, we have compared different techniques to measure the time lag\footnote{For a given light curve, a proper estimate of the noise associated to the time lag, is discussed in \cite{Alexander1997} and \cite{Peterson1998}.}: 1) we search for the maximum of the correlation function, assuming that the latter is positive and smaller than 1000 days; 2) we fit a Gaussian to the main peak of the correlation function; 3) we measure the centroid of the correlation function. In the last two cases, we search for the peak after an automatic identification the main peak, assumed to be the signal located between the two minima of the correlation function for $\tau \in [-200, +400]\,$days. Figure~\ref{fig:DCFreal} displays the distribution of lags derived using those three methods for the ICCF and the DCF. The distribution of lags of the ICCF is well described by a Gaussian with a width of $\sim 12\,$days, centered on the input lag. Hence, the degradation of the observing conditions mostly introduces noise, but does not bias the time-lag measurement if the ICCF is used. This is not true when we use the DCF. In that case the distribution becomes broader and multi-modal, and the lag can either be biased low or high depending of the method used to measure it. This behavior is caused by the secondary peak visible in the DCF at $\tau \sim $ 220 days. As suggested by Fig.~\ref{fig:DCFreal}, this peak seems to be associated to the gaps in the light curve but it is not obvious that its location can be predicted a priori. Indeed, this peak is only marginally detected in the DCF of the continuous light curve with gaps (blue band in Fig.~\ref{fig:DCFreal}) but for a smaller lag.


\begin{figure}
\includegraphics[scale=0.45]{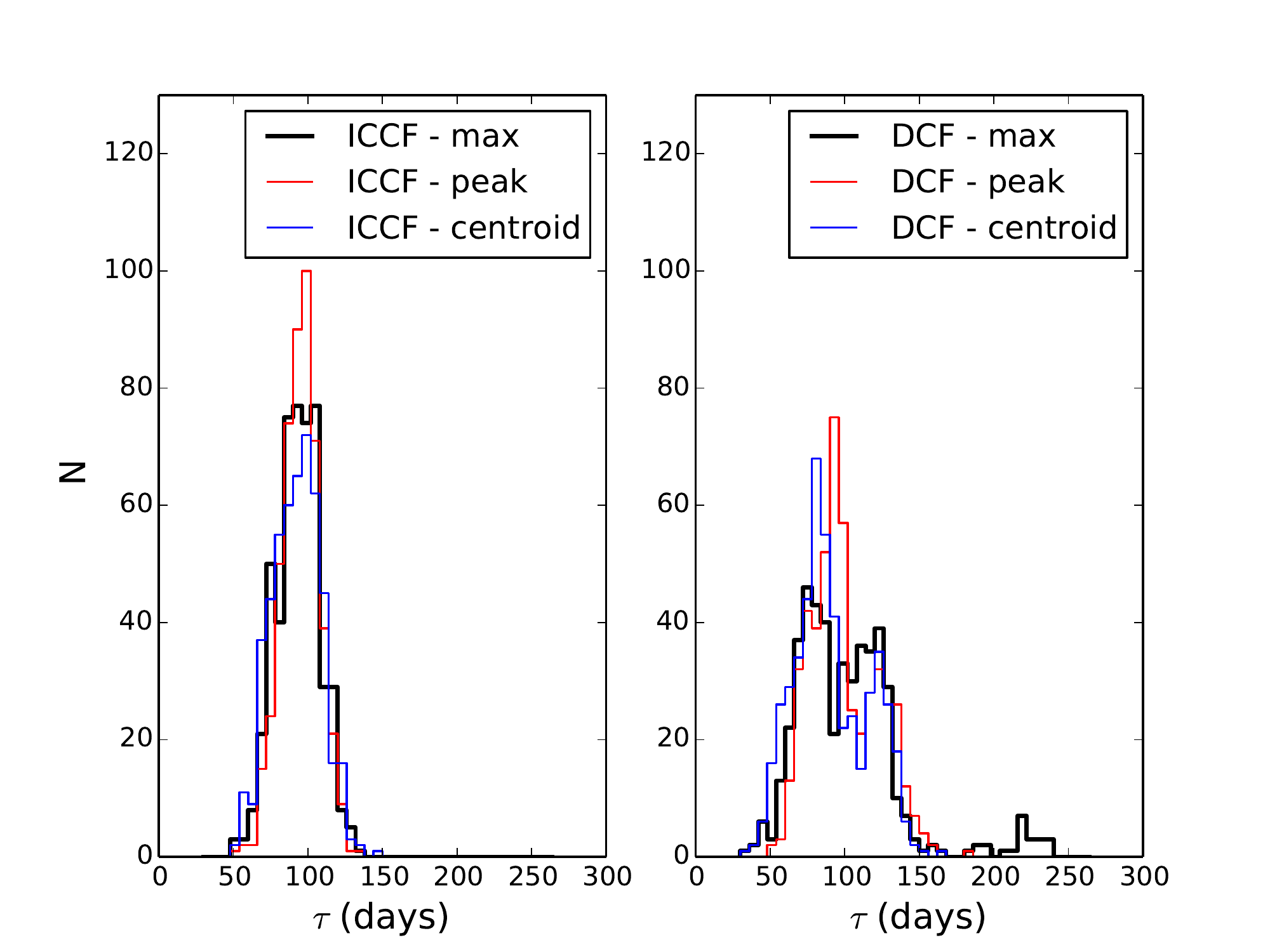}
\caption{Distribution of lags $\tau$ measured on the fiducial light curve for different realizations of the sampling and location of the gaps. The {\it{left}} panel shows the result for the ICCF and the right one for the DCF. The distributions are shown for three different estimators of the lag (cf Sect.~\ref{subsec:sampling}): 1) the maximum of the correlation function for lags shorter than 1000 days (thick black); 2) The maximum of a Gaussian function fitted on the main peak;  3) The centroid of the main peak.  }
\label{fig:DCFhist}
\end{figure}

\subsection{Time variable microlensing signal}
\label{subsec:micro}

Until now, we have assumed that $\mu$ is perfectly known and does not vary with time. Although this assumption may be valid in some realistic situations, the amplitude of microlensing generally varies on time scales of several months to several years \citep{Mosquera2011}. Rapid microlensing variations may be difficult to deblend from intrinsic variability when both variations occur on the same time scale, as observed in Q\,0158-4325 \citep[][]{Morgan2012a}. Such a situation is however rare and in most of the known lensed systems, the microlensing signal modulates the intrinsic variability with a lower frequency signal \citep[e.g.][]{Vuissoz2008, Hutsemekers2010}. In practice, extrinsic variability can often be modeled as a low order polynomial or a spline \citep{Kochanek2006, Tewes2013a}, corresponding to $\hat{F}_{12}(t)$ in (\ref{eq:A}). In the following we use spline models of $\hat{F}_{12}(t)$ and follow the prescription of Sect.~\ref{subsec:MmD} to derive $F_M$ and $F_{M\mu}$ with the MmD (Eq.~\ref{eq:MmD}). Then, we look whether the cross correlation of $F_M$ and $F_{M\mu}$ lead to an accurate estimate of $\tau$. 

\subsubsection{Generating the microlensing signal}

 We generate mock light curves similar to the fiducial one (i.e. $f_{BLR}=0.2$, $\tau_{fid}=100$\,days, $\sigma_f/f = 0.01$ and same intrinsic signal $F_M, F_{M\mu}$ as Sect.~\ref{sec:fiducial}), but now with time variable microlensing. The synthetic microlensing light curve is obtained by drawing 500 random trajectories in two different microlensing magnification maps constructed with the inverse ray-shooting code developed by \cite{Wambsganss1990, Wambsganss2001}. We used two maps representative of microlensing occurring in a saddle-point image with macro magnification $\mu\sim 20$ ($(\kappa, \gamma) = (0.47, 0.57)$), and a minimum image characterized by $\mu\sim 12$ ($(\kappa, \gamma) = (0.42, 0.50)$). This arbitrary choice, which simulates microlensing for the images A \& B-C of the lensed quasar RXS~J1131$-$1231, should be representative of microlensing in many lensed AGNs \citep[][ their Fig.~2]{Vernardos2014}. However, due to the large magnification of the images, the fraction of large amplitude microlensing events over a period of 10 years may be larger than commonly observed. In addition, we set the fraction of objects in compact form towards the lensed images to $f_* \sim 7\%$, typical of the stellar fraction at a galacto-centric distance of a few effective radii, i.e. where lensed images are commonly located \citep[][but see Jim\'enez-Vicente~\citeyear{Jimenez2014} who derived $f_* \sim 0.2$]{Mediavilla2009, Pooley2012}. We assume that the continuum arises from a disc with half-light radius $R_{1/2}=$0.06\,$\eta_0$ (where $\eta_0$ is the Einstein radius of a microlens), and a track length of 0.45\,$\eta_0$, which corresponds to a transverse velocity of 0.05\,$\eta_0/{\rm{year}}$. Those estimates match expectations for known lensed quasars ~\citep{Mosquera2011}.

\subsubsection{Microlensing models and time-lag measurement}
\label{subsec:lag}

\begin{figure*}
\centering
\includegraphics[scale=0.5]{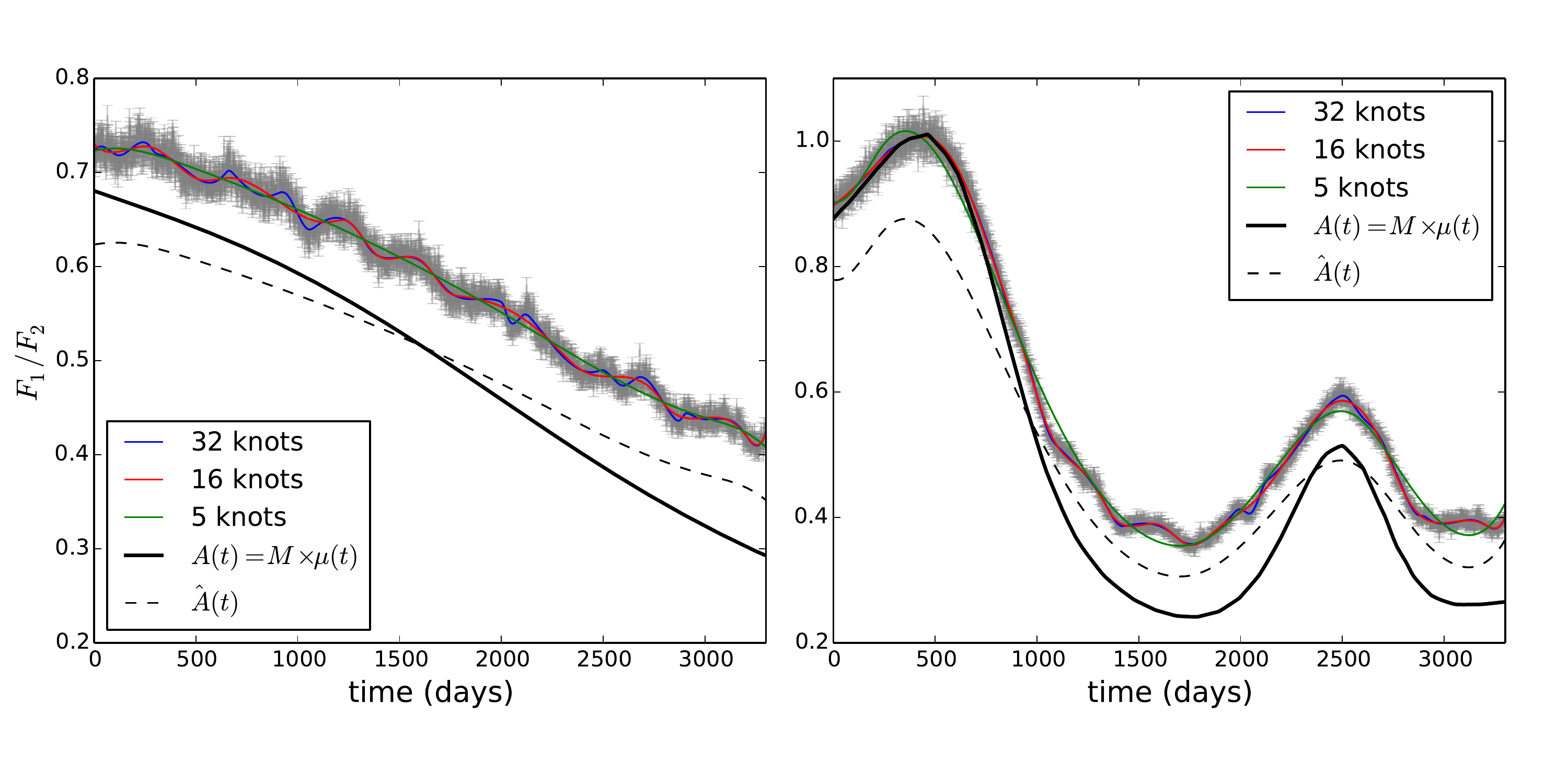}
\caption{Two examples of differential light curves and associated spline models $\hat{F}_{12}$. The grey points represent the simulated flux ratio with associated error bars. The thin solid lines show the spline models. The thick solid black curve shows the true microlensing variability $A(t)$ and the dashed one shows $\hat{A}(t)$ (Eq.~\ref{eq:A}) when $\hat{F}_{12}$ is a spline with 5 knots.}
\label{fig:splines}
\end{figure*}

We use the \texttt{PyCS} package \citep{Tewes2013a} to construct, from the pair of simulated light curves, an empirical model of the variability \citep[see][ Sect. 4]{Tewes2013a}. The intrinsic variability signal, common to the pair of light curves, and the extrinsic signal, are simultaneously fitted to the data with separate {\it {free-knot}} spline models\footnote{The total number of knots used for the spline is fixed by the user but the position of these knots is free.}. In addition to the parameters of the spline model, the magnitude shift between the curves and the time delay, are free to vary. We employ that technique because it is now commonly used for time-delay measurements \citep{Courbin2011, Tewes2013, Eulaers2013, Kumar2013}. It provides very good fit to our synthetic light curves, for which we also retrieve our input delay of 0 days. Figure~\ref{fig:splines} displays various splines reproducing the differential light curve $F_1/F_2$, and compares this signal to the input microlensing signal. We compare three splines for the extrinsic variations, differing by their number of ``free knots''. The spline with the largest number of knots (32) reproduces best the extrinsic variations but includes variations which are not those of the microlensed continuum. With a lower number of knots (5), the modeling of the light curves is poorer but the model better represents the microlensing variations. \\

\begin{figure*}
\centering
\includegraphics[scale=0.5]{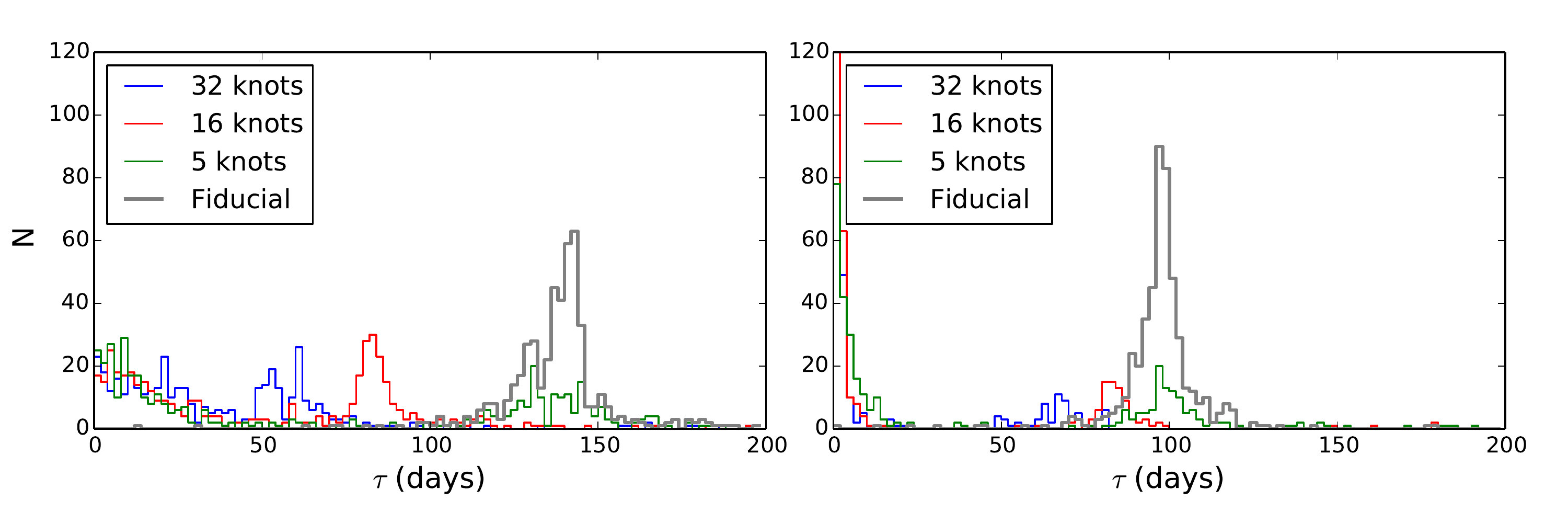}
\caption{Distribution of lags $\tau$ measured on the fiducial light curve for different realizations of time variable microlensing  (cf Sect.~\ref{subsec:micro}). {\it{Left:}} Distribution of lags $\tau$ using $\hat{F}_{12}(t)=A(t)$ (Eq.~\ref{eq:A}).  {\it{Right:}} Distribution of lags using  $\hat{A}(t)$, and assuming that the lag $\tau<210\,$days as suggested by the left panel. Each panel shows the result for three different spline models of the extrinsic variability, and applying the same procedure to the fiducial case where microlensing is constant over time.}
\label{fig:microvar}
\end{figure*}

Figure~\ref{fig:microvar} shows the distribution of time lags derived with the three spline models described above. The results for the fiducial light curve, namely when microlensing does not vary with time, are also shown. On the left panel, we show the lag as measured using $\hat{F}_{12}(t)$ in the MmD, while the right panel assumes that $\hat{A}(t)$ (Eq.~\ref{eq:A}) has been used. If we focus on the fiducial case, we see that the distribution of lags peaks at $\tau \sim 150$\,days when $\hat{F}_{12}(t)$ is used. This bias is expected because  $F_M$ contains a fraction of the microlensed flux. Once we use $\hat{A}(t)$, we recover the true lag for the fiducial distribution. The latter is the reference towards which lags derived in presence of time variable microlensing have to be compared. The spline model with 5 knots leads to a distribution of measured lags similar to the fiducial one, while more flexible spline models remove a fraction of the intrinsic signal and bias, or even preclude, the lag measurement. When using $\hat{F}_{12}(t)$ in the MmD, we find for a significant fraction{\footnote{For the chosen magnification maps, a wrongly estimated lag is derived for about 50\% of the curves. This fraction drops to 5\% if the distribution of microlensing event shows smaller amplitude variations as it is the case when e.g. the fraction of compact microlenses $f_*=1$.}} of the light curves lags with  $\tau < 50\,$days or $\tau > 200\,$days. Those incorrect lags are derived in two cases: when the microlensing signal varies so quickly that its variations are not adequately reproduced by the spline model, and when both magnification and demagnification occur during the monitoring. In the latter case, the MmD fails once $\hat{A}(t) \sim M$ (i.e. $\mu \sim 1$). The use of the DCF instead of the CCF, reduces the weight of those regions in the correlation function, but generally still produces a peak at $\tau \sim 0\,$ days, or occasionally at $\tau > 200 \,$ days. The correct time lag can be recovered for most of those light curves if the lag at $\tau \sim 0\,$ days is ignored or if the error bars on $F_M-F_{M\mu}$ are artificially increased for the time range where $A(t) \sim (1\pm 0.3) M $. Similarly, increasing the number of knots in the spline allows the recovery of the time lag for most of the light curves where $F_1/F_2$ is poorly approximated by a 5-knots spline. When we use $\hat{A}(t)$, we find a distribution of lags compatible with the fiducial one. Although this procedure properly shifts the first peak of the DCF to the correct lag, it often increases the height of the second peak at $\tau > 500\,$days. Therefore, our automatic identification of the peak of the DCF sometimes fails to identify the peak at $\tau \sim 100\,$days, artificially decreasing the number of retrieved lags. A visual inspection of the DCF alleviates this problem.  

The distribution of recovered lags does not perfectly match the fiducial one, suggesting that a bias at a few percent level might exist in the lag measurement. In addition, we find that time lags are more likely to be correctly retrieved when the average amplitude of microlensing over the monitoring period is larger than typically 0.75 mag. Among all events, time lags are more efficiently measured when the continuum is strongly demagnified. 

We have currently assumed that the amplitude of microlensing of the BLR is negligible. However, it has been shown that 10-20\% of the flux of broad line is typically microlensed \citep{Sluse2012b, Guerras2013}. To test for this effect, we have assumed that the whole BLR responds to the flux from the continuum, and is microlensed according to its half-light radius size $R^{BLR}_{1/2}$. We have chosen $R^{BLR}_{1/2} \sim 3\eta_0$, typical of the size of the BLR in known lensed quasars \citep[][their Fig.~3]{Mosquera2011}. Because the BLR size is in units of Einstein radius (and therefore depends on the source and lens redshifts) we do not scale the time lag according to the BLR size. Considering only reliable lags (i.e. such that $\tau \in [50, 200]$\,days), we find a distribution of lags in statistical agreement with those measured for a non-microlensed BLR. 

\subsubsection{Summary}

The above results demonstrate that it is possible to correct empirically for the time variable microlensing signal without removing the imprinted time-lagged signal from the BLR. As expected, time lags are more easily retrieved when microlensing evolves almost linearly over the period of monitoring, as often observed in lensed quasars light curves \citep{Vuissoz2007, Vuissoz2008, Courbin2011, Eulaers2013}. When the microlensing variations are large, some fine tuning may be necessary to detrend the light curves with a spline model. On one hand, the spline used to model extrinsic variations should not be too flexible, since it can then remove signal which is not associated to differential microlensing. On the other hand, regions of the light curves where there is no differential microlensing (i.e. $\mu \sim 1$) lead to an artificial peak at $\tau \sim 0\,$ days in the DCF. Ignoring this peak, or artificially increasing the error bars of the concerned points in $F_M$ and $F_{M\mu}$, often solves the problem and allows one to identify the peak produced by the lagged BLR signal.  

\section{Impact on time-delay measurements}
\label{sec:dt}

The presence of an intrinsic variability signal in the difference light curve between pairs of lensed images could be a source of systematic errors for the measurement of the (strong-lensing) time delay between the light curves. To test for this effect, we use \texttt{PyCS} to measure time delays of a set of mock light curves. The measurement of the time delay is a complex problem by itself \citep[see, e.g.,][]{Eigenbrod2005, Tewes2013a}, which depends on the length of the light curve, on the shape and amplitude of the variability signal, on the time sampling, and presence of observing gaps. Therefore, we will not quantify biases possibly taking place depending of these properties but limit ourselves to simple and well controlled situations. 

Our intrinsic signal is the same as the fiducial one, with $f_{BLR}=0.2$ and $\tau=100$\,days and Gaussian noise with $\sigma_f/f = 0.01$. The light curve is shortened to 3099 days, and the observing rate is chosen to be one point every three days on average. Two groups of light curves, one with a slowly varying microlensing signal and one with a large amplitude microlensing, as depicted in Fig.~\ref{fig:splines}, are generated. For each group, three different delays $\Delta t = 20, 100$ and $200$\,days, shorter, similar or larger than the time lag $\tau$, are assumed. We hope this ensemble of light curves to be sufficiently representative of real light curves. For each situation, we generate 500 light curves which have different sampling and noise realizations. 

To measure the time delay, we follow the prescriptions of \cite{Tewes2013a}, and fit a sum of two free-knot spline models, one for the intrinsic and one for the extrinsic light curve (cf. Sect.~\ref{subsec:micro}). In addition, we also apply the same technique to a set of reference light curves with the same characteristics as above but containing only flux from the continuum (i.e. there is no flux from the BLR). Following that procedure, we find that time-delay measurements are unaffected by the lagged signal from the BLR, provided the spline used to model the extrinsic variability is flexible enough. A comparison of the three spline models shown in Fig.~\ref{fig:splines} reveals that a small bias of a fraction of a day may take place when the spline used for microlensing has few knots. This suggests that methods employing an insufficiently flexible microlensing model could suffer from biases due to the quasar structure, in addition to the biases due to the poorly fitted microlensing variability. A detailed comparison of time-delay measuring techniques in presence of quasar structure is beyond the scope of the present paper. 

\section{Summary and conclusions}
\label{sec:conclusions}

Nowadays, high quality photometric monitoring data of gravitationally lensed quasars are obtained for more then 30 systems. In the next decades, owing to the advent of large surveys like the LSST, the time domain will be accessible for an increasing number of astrophysical phenomena, and in particular for quasar and gravitational lensing studies. Current analysis of optical lensed quasars light curves implicitly assume that the flux originates only from the accretion disc. Under this assumption, the differential signal between pairs of light curves shifted by the time delay, yields to the time variable microlensing signal produced by the stars in the lensing galaxy. However, the hypothesis that the entire broad band flux originates from a single emitting region does not hold in general. Broad-band quasar emission includes flux from the broad emission lines, and to a lesser extent from narrow emission lines, and other sources of emission (e.g. Balmer continuum, flux from the host galaxy). Because the various emitting regions have different sizes, they are affected differently by microlensing. The latter modifies the contrast between the continuum and the other sources of emission, in particular the broad lines (Fig.~\ref{fig:fidu}). Since broad lines respond to continuum variations, their light curve can be cross correlated to the one of the continuum to derive their size. This is the exactly what is performed by the reverberation mapping technique, which is one of our most powerful probe of the structure of the broad line region (BLR), and a robust proxy to the mass of the central black hole. That technique, originally designed for spectrophotometric data \citep[e.g.][]{Peterson1993, Kaspi2000, Bentz2009, Denney2010, Pancoast2011}, is currently extended to multi-band photometric data \citep[e.g.][]{Chelouche2012, Zu2014}. The signal produced by microlensing on quasar light curves, is conceptually very similar to photometric reverberation mapping, but is potentially applicable to single band data. In addition, since microlensing provides an independent probe of the accretion disc and BLR size \citep{Kochanek2004a, Eigenbrod2008b, Morgan2010a, Blackburne2011a, Sluse2011a, Guerras2013}, it offers a potentially more complete picture of the same sample of objects. 

We have studied the modulation of the differential microlensing signal between pairs of lensed quasar light curves when broad band emission originates from two regions: the accretion disc and the BLR. Assuming intrinsic quasar variations of about $0.5$ mag, and $\sim$20\% of the emission originating from the broad line, we have shown that modulation of the microlensing signal as large as $0.05$\,mag could be detected for differential microlensing larger than $0.5$\,mag. We have introduced a technique to combine pairs of lensed quasars light curves which enables one to disentangle the flux of the continuum and of the broad line, provided a spectra of the lensed images have been obtained once during the monitoring to derive the macro-lensing flux ratio. The measurement of the time lag $\tau$ between the continuum and BLR variations obtained this way, is what we named ``microlenisng aided reverberation mapping''. This technique has been applied to several sets of mock light curves, under the simplifying hypothesis of constant microlensing, in order to test the ability of the method to retrieve $\tau$ under various observational conditions. We found that unbiased time lags could be retrieved for fractional flux from the BLR as small as $\sim 10\%$ of the continuum flux, and amplitude of microlensing as small as $0.25$ mag. The precision on the time-lag measurement depends little on the fraction of flux from the BLR (provided it is typically $> 10\%$), but requires sufficiently large amplitude of microlensing, and photometric uncertainties typically better than $0.04$\,mag, to measure a time lag with small uncertainties. The precision depends significantly on the rate at which data points are observed, and more crucially on the absence/presence of seasonal observing gaps. Gaps and sparse sampling might bias the measurement of $\tau$ depending of the cross correlation technique used. We suggest that more advanced methods, such as the damped random walk models designed by \cite{Zu2011} combined with microlensing simulations, may offer a framework to tackle that problem. 

We have tested if a time-variable microlensing is a strong obstacle to time-lag measurements. We found the use of a spline to model the extrinsic variability to be efficient in detrending light curves from microlensing, but some fine tuning in the spline model is often necessary to obtain optimal results. More problematic have been the light curves showing both deamplification and amplification of the continuum during the observational period. In that case, the decomposition method we introduced fails to deblend the continuum and the BLR signal over the monitoring period, and leads to spurious peaks in the cross correlation. Ignoring the time range where there is no differential microlensing between the quasar images allows one to generally solve the problem and recover the correct lag. 

Finally, we have performed a preliminary investigation of the impact of the above discussed effect on the measurement of the time delay between lensed quasar images. Multiple techniques exist to measure time delays. We have focused on the one introduced by \cite{Tewes2013a} which uses a free-knots spline function to model the intrinsic and extrinsic variability of quasar light curves. This method, applied to mock light curves with time delays larger, equal or shorter than $\tau$, robustly retrieves the input delay provided the spline modeling the extrinsic variability is flexible enough. In such a situation, the spline function does not only model the microlensing, but also the intrinsic signal superimposed to it. We anticipate that methods which do not account for extrinsic variations in a sufficiently flexible way may lead to a biased estimate of the time delay. A time-delay challenge has been recently set up to test the ability of current techniques to measure accurate time delays from the thousands of lensed quasars light curves which should be monitored with LSST \citep{Dobler2013, Liao2014}. We suggest that future time-delay challenges account for the effect outlined in this paper. 

Microlensing-aided reverberation mapping is a promising technique to study the quasar structure from light curves of time-delay lensed quasars up to high redshift. The natural magnification of the lensed images which happens as a consequence of strong lensing, offers a natural boost of signal to noise. The small image separation, the time delay between the lensed images and the microlensing produced by the stars in the lensing galaxy were in the past a strong limitation in the use of lensed systems to study quasars. Current observational techniques and analysis methods allow one to tackle those difficulties, and use those systems as powerful astrophysical laboratories. While the analysis of the microlensing signal can be used to derive the size and temperature profile of the continuum emission \citep{Anguita2008a, Bate2008, Floyd2009a, Eigenbrod2008b, Poindexter2010a}, the intrinsic variability can be studied in the same systems to derive the size of the broad-line region. Although, we have conceptually demonstrated the feasibility of this technique, more exhaustive set of simulations need to be carried out in the future. For example, it is necessary to estimate the impact on time-lag measurements of subtle effects likely to take place in real data, such as contribution of non-variable flux (from an intermediate line region, from the narrow lines and/or from the host galaxy) to the broad-band signal. It has also to be seen how those results depend on the absolute value of the time lag, on the properties of the intrinsic variability signal, and on multiple observational constraints which might refrain us to model accurately the time-variable microlensing signal. 

\begin{acknowledgements}
DS acknowledges support from the Deutsche Forschungsgemeinschaft, reference SL172/1-1, and MT acknowledges support by the DFG grant Hi 1495/2-1. We thank the referee for his/her prompt and useful report. This research made use of Astropy, a community-developed core Python package for Astronomy \citep{Astropy2013}.
\end{acknowledgements}

\bibliographystyle{aa}
\bibliography{/home/sluse/work/articles/bibds}
\end{document}